\documentclass[journal]{IEEEtran}
\ifCLASSINFOpdf
\else
\fi
\usepackage{url}
\usepackage{import}
\usepackage[detect-all=true,separate-uncertainty=true, multi-part-units=single,range-phrase={-},product-units = single]{siunitx}
\usepackage{menukeys}
\usetikzlibrary{arrows,positioning,shapes.geometric}
\usetikzlibrary{intersections}
\usepackage{amsfonts}

\begin{document}
%
\title{Ultrasound Shear Wave Elasticity Imaging with Spatio-Temporal Deep Learning}                 
%
%
%

\author{Maximilian~Neidhardt*, 
Marcel~Bengs*, 
Sarah~Latus, 
Stefan~Gerlach,\\
Christian~J.~Cyron, 
Johanna~Sprenger 
and Alexander~Schlaefer
\thanks{M. Neidhardt, M. Bengs, S. Latus, S. Gerlach, J. Sprenger and A. Schlaefer are with the Institute
of Medical Technology and Intelligent Systems, Hamburg University of Technology,
Am-Schwarzenberg-Campus 3, 21073, Hamburg, Germany, e-mail: maximilian.neidhardt@tuhh.de.}
\thanks{C. J. Cyron is with the Department of Continuum and Materials Mechanics,
Eißendorfer Straße 42 (M), 21073, Hamburg, Germany \\e-mail: christian.cyron@tuhh.de.}
\thanks{This research was partially funded by the DFG SCHL 1844/2-2 project and by the TUHH i$^{3}$ initiative.}
\thanks{* Both authors contributed equally.}
\thanks{Copyright (c) 2021 IEEE. Personal use of this material is permitted. However, permission to use this material for any other purposes must be obtained from the IEEE by sending an email to pubs-permissions@ieee.org.}
}

\markboth{IEEE TRANSACTIONS ON BIOMEDICAL ENGINEERING,~Vol.~XX, No.~X, April~2022}%
{Neidhardt \MakeLowercase{\textit{et al.}}: US Elasticity Imaging with Deep Learning}
%



\hyphenation{time-of-flight}

\maketitle

\begin{abstract}
Ultrasound shear wave elasticity imaging is a valuable tool for quantifying the elastic properties of tissue. Typically, the shear wave velocity is derived and mapped to an elasticity value, which neglects information such as the shape of the propagating shear wave or push sequence characteristics. We present 3D spatio-temporal CNNs for fast local elasticity estimation from ultrasound data. This approach is based on retrieving elastic properties from shear wave propagation within small local regions. A large training data set is acquired with a robot from homogeneous gelatin phantoms ranging from \SI{17.42}{\kilo\pascal} to \SI{126.05}{\kilo\pascal} with various push locations. The results show that our approach can estimate elastic properties on a pixelwise basis with a mean absolute error of \SI{5.01(437)}{\kilo\pascal}. Furthermore, we estimate local elasticity independent of the push location and can even perform accurate estimates inside the push region. For phantoms with embedded inclusions, we report a 53.93\% lower MAE (\SI[]{7.50}{\kilo\pascal}) and on the background of 85.24\% (\SI[]{1.64}{\kilo\pascal}) compared to a conventional shear wave method. Overall, our method offers fast local estimations of elastic properties with small spatio-temporal window sizes.
\end{abstract}


%
\IEEEpeerreviewmaketitle

\section{Introduction}
\label{sec:introduction}
\IEEEPARstart{Q}{uantifying} mechanical properties of soft tissue has many clinical applications ranging from diagnoses \cite{cosgrove2013efsumb} to modeling soft tissue response for surgical planning \cite{umale2013experimental}. Ultrasound shear wave elasticity imaging (US-SWEI) is widely used to image the elastic properties of tissue and its clinical applications has been widely demonstrated, e.g., in disease staging of breast lesions \cite{yang2017qualitative}, thyroid nodules \cite{sebag2010shear} or liver fibrosis \cite{sande2017ultrasound}. In US-SWEI, an initial high energy acoustic radiation force impulse displaces the tissue. The propagation of the resulting shear wave is then captured with high frequency ultrasound imaging.

Shear wave velocity is commonly used as a surrogate for tissue elasticity, which can be estimated from a sequence of images, considering either the time-domain or the frequency domain. The first approach tracks the peak of the propagating shear wave, often referred to as time-of-flight (ToF). This can be achieved either by applying an autocorrelation of two time-varying signals with a known distance between each other (\cite{bercoff2004supersonic,song2014fast,latus2017,song2012comb}) or by performing a linear regression of the wave peaks in a 2D space-time image (\cite{carrascal2017improved, Engel2015}). Commonly, ToF-methods assume that shear waves propagate in a fixed direction. To estimate wave velocity independently of the propagating direction, 2D-autocorrelation methods were proposed (\cite{song2012comb, song2014fast}). In general, ToF-methods have been evaluated in the clinical setting \cite{tanter2008quantitative}. However, estimates are dependent on imaging depth \cite{yang2014comparative} and performance has been reported to be limited for stiffer materials \cite{wang2013precision}, which are characterized by faster shear waves. The second approach for US-SWEI estimates the phase velocity of the dominant local wavenumbers in the frequency domain \cite{kijanka2018local}. Similar to a 2D-autocorrelation, this approach is independent of the wave direction but requires intensive tuning of the imaging and filter parameters \cite{Kijanka2020}.

\begin{figure}[t!]
    \import{./}{fig_methodOverview.tex}
\end{figure}  

\begin{figure*}[!t]
    \centering
    \includegraphics[width=1.0\linewidth]{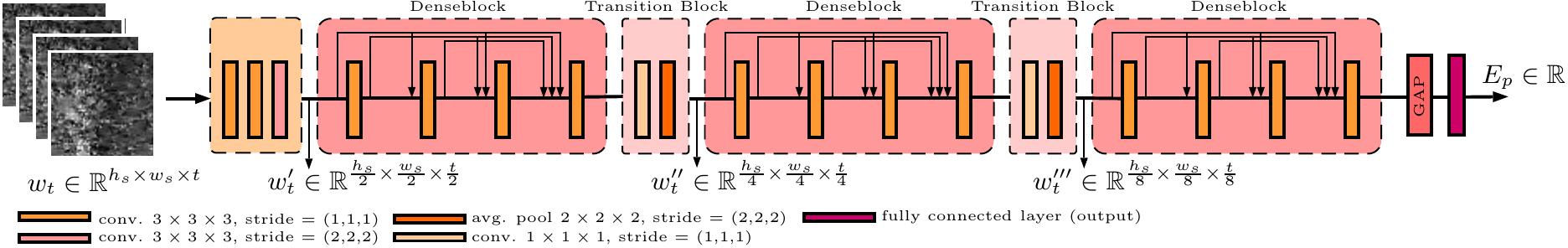}
    \caption{Spatio-temporal CNN architecture. We predict the Young's modulus from a 3D spatio-temporal window as input. Our network consists of initial convolutional layers, followed by DenseNet blocks. Between the DenseNet blocks we apply average pooling layers for downsampling of the input dimensions.}
    \label{fig:deep_learning_pipeline}
\end{figure*}

\textcolor{black}{Recently, deep learning methods have gained popularity in strain elastography (\cite{kibria2018gluenet, chan2021deep, kz2020semi, delaunay2021unsupervised}) and SWE-imaging (\cite{jin2021deep, vasconcelos2021viscoelastic,ahmed2021dswe}).} These methods allow estimates without intensive preprocessing of the data, manual tuning and do not rely on feature extraction, e.g., the shear wave velocity for elasticity estimation. Previous works have demonstrated that deep learning can be used to estimate distinct tissue parameters from SWEI data. Jin et al. \cite{jin2021deep} predict the shear wave velocity from space-time images including an uncertainty estimate and Vasconcelos et al. \cite{vasconcelos2021viscoelastic} have shown that the viscoelastic model parameters can be estimated from simulated shear wave motion data. Further, Ahmed et al. \cite{ahmed2021dswe} demonstrated that elasticity maps and segmentation masks can be generated with deep learning from simulated SWEI data. However, the authors also note that the use of simulated data does not seem to be sufficient to represent the noise in real data.

In this study, we present spatio-temporal convolutional neural networks (CNNs) for reconstructing elasticity maps from real ultrasound SWEI data. Our approach is based on the concept of retrieving local elastic properties from shear wave propagation in small regions of several millimeters, which we call spatio-temporal windows (Figure~\ref{fig:our_approach}). By performing this local elasticy estimation, the network learns the direct relationship between localized shear wave propagation and local elasticity. Our localized approach enables the generation of detailed elasticity maps of inhomogeneities and simplifies the required training data. In particular, this allows us to acquire training data from simple homogeneous phantoms with defined elasticities. Using real ultrasound training data, probe and ultrasound artifacts are directly included in our approach. In this context, we systematically study whether deep learning is able to extract relevant information from these limited areas and whether this approach generalizes to different push locations and elasticities. We evaluate our approach using tissue mimicking gelatin phantoms with Young's moduli ranging from \SI{17.42}{\kilo\pascal} to \SI{126.05}{\kilo\pascal}. Furthermore, we compare our spatio-temporal CNN approach to ToF shear wave estimation. 

In summary, our 3D spatio-temporal CNN approach can estimate local elastic properties from spatio-temporal windows using real ultrasound shear wave data, while being independent of the push location and wave propagation direction. Furthermore, our approach can generate local elasticity maps of non-homogeneous mediums in real-time.


\section{Methods}

\subsection{Deep Learning Model}
We estimate the elasticity locally by applying spatio-temporal CNNs to a small spatio-temporal window as illustrated in Figure~\ref{fig:our_approach}. Hence, we perform pixelwise predictions using the neighborhood as context. Formally, given a sequence of images $x\in \mathbb{R}^{h \times w \times t}$ which represent shear wave propagation over time with $h$ and $w$ for the spatial dimensions of the FOV and $t$ for the temporal dimension, the elasticity is estimated locally using a spatio-temporal window $\widetilde{x}\in \mathbb{R}^{h_{s} \times w_{s} \times t}, \widetilde{x} \subset x$ centered at pixel location $p=[p_{i},p_{j}]$. The spatial dimensions of the spatio-temporal window are described by $w_{s}$ and $h_{s}$. Hence, we design and evaluate an approach for learning $f\colon\,\mathbb{R}^{h_{s}\times w_{s}\times t}\rightarrow\mathbb{R}$. By using our spatio-temporal CNN to estimate elasticity for each pixel, an entire global elasticity map for $x$ can be estimated, as shown in Figure~\ref{fig:our_approach}. The advantage of local elasticity estimation is that the network is trained to learn the relationship between elasticity and shear-wave propagation only for a small spatial patch. In this way, the network can be trained with data from simple homogeneous phantoms, while also being applicable to inhomogeneous data subsequently.

Spatio-temporal CNNs \cite{tran2015learning} have demonstrated promising results for imaging elastic properties using optical coherence elastography (\cite{neidhardt2020deep,neidhardt20214d}). The concept is to apply convolutions jointly over space and time, which enables spatio-temporal feature learning from data \cite{tran2015learning}. In this way, local spatio-temporal dependencies, which are present for the spatio-temporal windows $\widetilde{x}$, are learned and extracted. As a baseline, we consider the concept of Densely Connected Convolutional Networks (DenseNet) \cite{huang2017densely} due to its parameter and computational efficiency and develop our own custom DenseNet architecture (\cite{neidhardt2020deep, neidhardt20214d}).  Our architecture details are shown in Figure~\ref{fig:deep_learning_pipeline}. Our 3D architecture consists of three initial convolutional layers, followed by three DenseNet blocks with four convolutional layers each. Between the DenseNet blocks, we apply average pooling layers for downsampling of the input dimensions. Moreover, we use batch normalization \cite{batchnorm2015} and the rectified linear activation function for our convolutional layers. Using this 3D CNN architecture, we consider spatio-temporal windows $\widetilde{x}$ with a size of $65\times65\times t$, $33\times33\times t$, $17\times17\times t$, $9\times9\times t$ and $5\times5\times t$ with $t=35$ frames. We set the spatial stride of our architecture in Figure  \ref{fig:deep_learning_pipeline} to one for spatio-temporal window sizes of $9\times9\times 35$ and $5\times5\times 35$.

 \subsection{Conventional Shear Wave Velocity Estimation}
To compare our spatio-temporal CNN we consider a ToF approach. To reduce speckle noise, we apply a 3D mean filter with a kernel size of \SI{5}{px} along all axis. Furthermore, we process our data with a directional filter in the frequency domain to reduce waves that propagate along the lateral off-axis and to limit high-frequency imaging noise \textcolor{black}{ \cite{song2012comb}}. For a distinct pixel, we estimate the ToF by performing an auto-correlation between the time varying signals measured at an equivalent distance to the distinct pixel along the lateral axis. Using a distance of 65 pixel, we apply a Tukey window on the two time signals, interpolate them by a factor of 10 and subsequently estimate the time delay by auto-correlation. We assign the estimated shear wave velocity to the pixel located at the center between two measurement points. We reject estimates which are not in the range of \SIrange{0.1}{10}{\metre\per\second}. Our data processing is similar to Song et al. \cite{song2012comb}. Please note that we perform data acquisition with a single push sequence and subsequent high frequency plane wave imaging with a single steering angle of 0$^\circ$. Following \cite{sarvazyan2013acoustic}, the shear wave velocity $c_{s}$  is mapped to the Young's modulus with the relation
\begin{equation}
    E_{ToF} = \alpha \cdot \rho \cdot 2(1+\nu) \cdot c_{s}^{2}
\end{equation} 
with the density $\rho=\SI{1000}{\kilo\gram\per\cubic\metre}$ and the Possion's ratio $\nu=0.5$. For a fair comparison to our deep learning approach, we introduce a scaling factor $\alpha$, to account for constant errors between our estimates and our ground truth Young's modulus labels estimated from indentation experiments. We estimate $\alpha = 0.75$ by minimizing the offset between the mean of all Young's moduli estimates from a single gelatin concentration and the corresponding indentation experiment. 
For inclusion map experiments, we apply a gaussian filter with a kernel size of \SI{2}{} $\times$ \SI[]{2}{\milli\metre} to account for outliers and close holes with no predicted values due to noise. Also, we increase the FOV and average estimates by combining nine push and imaging sequences distributed evenly across the probe length.

 \begin{figure}[]
    \import{./}{fig_indentationExperiment.tex}
\end{figure}

\begin{table}[]
    \centering
\caption{Reported Young's moduli of soft tissues in the literature, estimated either by indentation [I] or shear wave elasticity imaging [SWEI]. } 
\label{tab:E_Modulus_Literature}%
\begin{tabular}{lccl}
Tissue                      & E [\SI[]{}{\kilo\pascal}]          & Method    & Reference \tabularnewline
\hline
Adipose tissue & $5.60$ & [I] &  \cite{Samani2003} \\
    Thyroid                     & $10.97\pm3.1$     & [SWEI]    &  \cite{arda2011quantitative}\\
    Procine Liver         & 10-15 & [I] &  \cite{samur2007robotic}\\
            Muscle                      & 12-13             & [SWEI]    &  \cite{kot2012elastic}\\
Spleen                      & 14-35             & [I]       &  \cite{umale2013experimental} \\
Renal Cortex                & $15\pm7.2$        & [I]       &  \cite{umale2013experimental} \\
   Prostate (Healthy)        &  $17.0\pm9.0$ & [I]  &  \cite{ahn2010mechanical}\\
       Normal Fat                  & 18-24             & [I]       &  \cite{krouskop1998elastic}\\
           Renal pelvis                & $23.60\pm5.4$     & [SWEI]    &  \cite{arda2011quantitative}\\
Prostate (Cancerous)        & $24.1\pm14.5$               & [I]      &  \cite{ahn2010mechanical}\\
 High Grade Brest Tissue     & $43\pm12$         & [SWEI]    &  \cite{samani2007elastic} \\
Achiles Tendon              & $51.50\pm25.1$     & [SWEI]   &  \cite{arda2011quantitative}\\
    Tendon                      & 70-72            & [SWEI]    &  \cite{kot2012elastic}\\
Fibrous Breast Tissue       & 96-244            & [I]       & \cite{krouskop1998elastic}\\

\hline 
\end{tabular}

\end{table}

 \begin{figure}[]
    \import{./}{fig_stressStrain.tex}
\end{figure}

\subsection{Phantom Preparation and Annotation}
We use gelatin phantoms as tissue surrogates and prepare batches of gelatin with a weight ratio of gelatin to water ranging from \SI{5}{\%} to \SI{17.5}{\%} in increments of \SI{2.5}{\%}. For precise and reproducible manufacturing, we thoroughly follow this procedure: mix ballistic gelatin (250 Bloom Type A Ordenance Gelatin, Gelita) and water, let the mixture mature for 2 hours, heat the mixture automatically controlled to \SI{50}{\degreeCelsius} and add \SI{1}{\gram} of graphite per \SI{800}{\gram} weight for ultrasound speckle. Experiments are performed after approximately 24 hours of cooling. Three types of phantoms are manufactured in-house: (1) for ground truth annotation we prepare eight cylindrical phantoms of each concentration with a radius \(\mathrm{r}=\SI{10}{\milli\metre}\) and a height \(\mathrm{l_{0}}=\SI{40}{\milli\metre}\) as shown in Figure~\ref{fig:IndentationSetup}, right, (2) for training and testing of our network we prepare block phantoms ($\sim$ \SI[]{100}{} $\times$ \SI{100}{} $\times$ \SI[]{100}{\milli\metre}) of each concentration and (3) inclusion phantoms with a gelatin concentration of \SI{7.5}{\%} for the background and a gelatin concentration of \SI{15}{\%} for embedded circular inclusions with a radius of approximately \SI{5}{\milli\metre} and  \SI{10}{\milli\metre} \textcolor{black}{, as well as embedded chicken heart tissue. To avoid gelatine layers, the casted cylindrical inclusions were fixed on both ends to the phantom wall before gelatin was added.}

We estimate the ground truth elasticity using the cylindrical phantoms. We perform unconfined compression tests to estimate the Young's modulus as the ratio of stress to strain (\cite{forte2015modelling, delaine2016experimental}). The experimental setup is shown in Figure~\ref{fig:IndentationSetup}. During indentation, the sensor records forces $F$ with a frequency of \SI{200}{Hz} and a resolution of \SI{3.1}{\milli\newton}. We restrict the applied forces to a maximum of \SI{2}{\newton}, drive the plate with a constant velocity of \SI{0.01}{\mm\per\s} and apply lubricant between contact surfaces to reduce bulking of the phantom. Viscous effects are reduced due to our slow indentation speed. To estimate the elasticity of our gelatin phantoms, we perform a single indentation experiment per phantom, to avoid material defects due to indentation. We perform 8 indentation experiments per gelatin concentration and consider the Young's modulus as 

\begin{equation}
    E=\frac{\sigma}{\epsilon}=\frac{F}{\pi r^{2}} \frac{l_{0}}{\Delta l}
    \label{Emodulus}
\end{equation}

with the stress $\sigma$ and strain $\epsilon$. We estimate $E$ with a linear regression applied to all indentation experiments performed on a gelatin concentration. The strain range is limited between \SI{2}{\%} and \SI{7}{\%}  (\cite{forte2015modelling, delaine2016experimental, zmudzinska2018assessment}). Results for the stress-strain curves and the corresponding Young's moduli are shown in Figure~\ref{fig:Stess_Strain_curve}. For comparison, elasticities of real tissue reported in the literature are presented in Table~\ref{tab:E_Modulus_Literature}.

\begin{figure}[]
    \import{./}{fig_dataAcquisition.tex}
\end{figure}

\subsection{Data Acquisition}
\label{sec:Ultrasound_Shear_Wave_Data_Acquisition}
Our experimental setup for US-SWEI data acquisition is shown in Figure~\ref{fig:Setup}. For pushing and imaging, we use a linear array probe (128 elements, 0.29 mm pitch, center frequency \SI{7.5}{MHz}) and a 128-channel ultrasound system (Cicada, Cephasonics Inc, USA). The ultrasound probe is positioned by a serial robot (UR3, Universal Robots, Denmark) for automatic data acquisition. A force sensor (Nano43, ATI, USA) is mounted to the end-effector of the robot to ensure a repeatable contact pressure of \SI{0.2}{\newton} between phantom and probe. Ultrasound gel is applied to the surface to reduce imaging artifacts. Once the transducer is positioned, an unfocused push sequence (\SI{120}{V}, 2000 push cycles, \SI{10}{\milli\metre} depth) excites a shear wave inside the phantom. \textcolor{black}{A continuous segment of 11 elements with the center element defined as the push location was used to transmit the unfocused push.} Subsequently, we perform plane wave imaging with an imaging frequency of \SI{7000}{\hertz} and a FOV of \SI[]{20}{} $\times$ \SI[]{33}{\milli\metre} along the depth and lateral image axis. We record 35 subsequent images after each push sequence with a resolution of \SI{250}{} $\times$ \SI[]{600}{pixels} along the depth and lateral axis, respectively, after beamforming. Image recording is engaged \SI{0.13}{\ms} after the push sequence to reduce imaging artifacts. Loupas's algorithm \cite{loupas1995axial} is applied on the IQ demodulated data to estimate axial displacement relative to a reference frame prior to excitation. We use the resulting displacement data as input to our network. In total, the robot positions the ultrasound transducer randomly at 80 positions on the surface of each gelatin block for data variation. At each position, we perform seven push and imaging sequences with individual pushes applied at the locations depicted in Figure~\ref{fig:Setup}, right. Using a robot allows us to efficiently acquire real ultrasound training data that includes probe and system characteristics.

\subsection{Training and Evaluation}
We train our approach with spatio-temporal windows $\widetilde{x}$ spatially located within a defined ROI with a size of $121 \times 181$ pixel ($10 \times 12$ mm), see Figure~\ref{fig:Setup}. To study the flexibility of our method \textcolor{black}{with respect to} push locations, we consider seven different push positions relative to the ROI. We train our networks using homogeneous phantoms with defined ground truth elasticity $E_{gt}$, determined by indentation experiments. Hence, we assign the corresponding ground truth elasticity $E_{gt}$ to a spatio-temporal window $\widetilde{x}$ and the learning task is to perform a regression of $\widetilde{x}$ to the corresponding elasticity $E_{gt}$. Therefore, a network is trained to learn the relationship between elasticity and shear-wave propagation for a small local region. For training, we minimize the mean squared error (MSE) loss function between the defined target ground truth elasticity $E_{gt}$ and our predicted elasticity $E_{p}$ defined as
\begin{equation}
    \mathcal{L}(E_{gt},E_{p})=\frac{1}{N}\sum_{k=1}^{N}\left\Vert  E_{gt}^{\{k\}}-E_{p}^{\{k\}}\right\Vert ^{2}
\end{equation}
 with $N$ for the number of samples. During one training epoch, we take one spatio-temporal window $\widetilde{x}$ with random location within the ROI from every image sequence $x$ in our training data set. Each network is trained for 250 epochs with a batch size of 250 using Adam for optimization with a learning rate of $l_{r}= 1e^{-4}$. After 150 epochs, we divide the learning rate by a factor of two every 50 epochs. We normalize the pixel intensities of each input $\widetilde{x}$ to have a zero mean and standard deviation of one. To augment our training data, we randomly apply horizontal and vertical flipping, multiple \ang{90} rotations, Gaussian blur and randomized input erasing of the input data during training.
 
We evaluate the performance of our method with all elasticities present during training and perform four-fold cross-validation on the 80 different positions of each concentration. For each fold we use 60 positions of each concentration for training, and 10 positions each for validation and testing. Second, we evaluate the regression performance of our method on unseen elasticities and perform a cross-validation approach, where we leave out the entire data of one gelatin concentration. We do not perform cross-validation on boundary elasticities, e.g., 5\% and 17.5\% as this leads to out of distribution predictions for the regression task. Hence, we perform four-fold cross-validation using the gelatin concentrations starting from 7.5\% up to 15\%. In each fold, we randomly split the data into 50\% of the fold's data for testing and 50\% for validation. Moreover, for all our trainings we remove push one and seven completely from our training data, to evaluate unseen push locations further away from the ROI. For elasticity estimation on inclusion phantoms, we refine the network trained on homogenoues phantoms, by fine-tuning the network for additional 10 epochs with inhomogeneous phantom data. Thereby, the network learns wave reflections at boundaries which are not present in homogenous phantoms.
 
 \begin{figure*}
    \import{./}{fig_singelPushPrediction.tex}
\end{figure*}

\begin{figure}
    \import{./}{fig_boxPlot.tex}
\end{figure}

\section{Results}
\begin{figure*}
    \import{./}{fig_pushLocation.tex}
\end{figure*}

\subsection{Homogeneous Phantoms}
We study our spatio-temporal CNN approach qualitatively to ToF in Figure~\ref{fig:singelPushPrediction} and evaluate the prediction maps of both approaches \textcolor{black}{with respect to} varying push locations and phantom elasticities. For this and the following evaluations, if not indicated otherwise, we consider the more challenging case for our deep learning approach where we left out entire elasticities during training. Our findings in Figure~\ref{fig:singelPushPrediction} demonstrate that our spatio-temporal CNN approach leads to more consistent estimations for all experiments. Notably, our results show that our spatio-temporal CNN provides estimations inside the push location in contrast to ToF that fails in general closer to the push region. Moreover, our results in Figure~\ref{fig:boxplot_push} confirm quantitatively that the performance of the spatio-temporal CNN is independent of the push location considering all phantom elasticities. Further, we study the performance of spatio-temporal CNNs and the ToF approach and evaluate the performance quantitatively \textcolor{black}{with respect to} phantom elasticity, see Figure~\ref{fig:boxplot_push}. Our results show that the pixelwise mean absolute error (MAE) increases with increasing elasticity for both ToF and spatio-temporal CNN. Our spatio-temporal CNN approach leads to an overall MAE of \SI{5.01(437)}{\kilo\pascal} when all elasticities are present during training, compared to an overall MAE of \SI{9.99(749)}{\kilo\pascal} where evaluated elasticities are left out during training. The ToF approach leads to an overall MAE of \SI{11.61(876)}{\kilo\pascal}.
Second, performance \textcolor{black}{with respect to} the spatio-temporal window size is given in Table~\ref{tab:All-networks-with metrics}. Our results demonstrate that larger spatial input sizes work better at the expense of reduced model throughput, e.g., using the largest spatial input sizes of \SI{65}{} $\times$ \SI{65}{pixels} ($\sim$ \SI{4}{} $\times$ \SI{5}{\milli\metre}) improves performance by 30\% while reducing the throughput by a factor of 31 compared to the smallest input size of \SI{5}{} $\times$ \SI{5}{pixels} ($\sim$ \SI{0.32}{} $\times$ \SI{0.4}{\milli\metre}). 

Third, we further study the robustness of our methods and show the standard deviation of predictions at each pixel for the complete FOV for the spatio-temporal CNN and the ToF approach using push one, four and seven, see Figure~\ref{fig:pushlocation}. Note that push one and seven are completely removed from the training data and that we only consider the ROI during training. Our results demonstrate that the spatio-temporal CNN provides consistent estimates with a low standard deviation also at the previously unseen push locations and at a larger FOV than the ROI. Moreover, Figure~\ref{fig:pushlocation} demonstrates that for lower phantom elasticity (\SI{37.55}{\kilo\pascal}) the predictions of the spatio-temporal CNN show a high standard deviation far away from the push location, similar to ToF. 

\subsection{Inclusion Phantoms}
Results for estimates using our spatio-temporal CNNs on phantoms with embedded cylindrical inclusions are shown in Figure~\ref{Fig_Inclusions} for spatio-temporal window sizes of 17$\times$17, 33$\times$33 and 65$\times$65 pixels. Depicted is the mean of nine push and imaging sequences. \textcolor{black}{We report the MAE for the phantoms backgrounds and inclusions separately by calculating the pixel-wise errors between the prediction of the network and the corresponding ground truth elasticity. The results for a spatio-temporal window size of 65$\times$65 pixels are given in Table~\ref{tab:InclusionMetrics}.} The combined MAE across all phantoms with deep learning is \SI[]{7.5(1287)}{\kilo\pascal} for all inclusions and \SI[]{1.64(0432)}{\kilo\pascal} for the background. The combined MAE with ToF is \SI[]{16.28(1005)}{\kilo\pascal} for all inclusions and \SI[]{11.11(1008)}{\kilo\pascal} for the background. The threshold for the Dice coefficient is set to \SI[]{67.38}{\kilo\pascal} and is estimated by the mean target Young's modulus of inclusion and background. The mean Dice coefficient for the inclusion shapes \textcolor{black}{depicted in Figure~\ref{Fig_Inclusions}} is 0.93 for our deep learning approach and 0.86 for ToF. Table~\ref{tab:inclusionMetricsWindowSizes} shows that the Dice coefficient and the MAE decrease for smaller spatio-temporal windows sizes. This is consistent for all \textcolor{black}{binarization thresholds} as shown in Figure~\ref{fig:diceWindowSizes}. \textcolor{black}{The elasticity map of chicken heart tissue, B-Mode ultrasound image and cross-section of the phantom is given in Figure~\ref{Fig_InclusionChicken}.}

 \begin{table}
    \caption{MAE and Pearson correlation coefficient (pCC) for different window sizes. Throughput refers to the number of positions for which elasticity can be estimated within one second. We measure the throughput of our methods on a NVIDIA Tesla V100-SXM2-32GB using a batch size of 500.} 
\label{tab:All-networks-with metrics}
\centering
\begin{tabular}{lccc}
    $h_{s} \times w_{s}$ & MAE [\SI[]{}{\kilo\pascal}] & pCC [\%] & Throughput [\SI[]{}{px\per\second}] \tabularnewline
    \hline 
    $5\times 5$     & \SI[]{14.40(1082)}{}      & \SI[]{72.81}{}    &   15745   \tabularnewline
    $9\times 9$     & \SI[]{12.63(0901)}{}      & \SI[]{86.53}{}    &   5651    \tabularnewline
    $17\times 17$   & \SI[]{11.83(0825)}{}      & \SI[]{87.54}{}    &  2348     \tabularnewline
    $33\times 33$   & \SI[]{10.64(0748)}{}      & \SI[]{88.55}{}    &   1688    \tabularnewline
    $65\times 65$   & \SI[]{09.99(0749)}{}      & \SI[]{93.39}{}    & 508       \tabularnewline
\end{tabular}
\end{table}

\section{Discussion}
We present a deep learning approach for local elasticity estimation from real 3D ultrasound data. This task has been addressed with conventional methods by extracting the shear wave velocity as an explicit feature (\cite{carrascal2017improved,Trutna2020}). In contrast, deep learning methods allow estimates without explicit feature extraction, intensive pre-processing and manual tuning. We present a local elasticity estimation approach with real ultrasound data, where a 3D spatio-temporal CNN is trained to predict tissue elasticity from spatio-temporal windows.

\textbf{Elasticity estimation.} We study the performance of our methods on homogeneous phantoms considering various push locations and elasticities. We demonstrate that predictions can be performed for an elasticity range of \SIrange{38}{98}{\kilo\pascal} which reflects reported tissue elasticities in the literature as shown in Table~\ref{tab:E_Modulus_Literature}. Our findings highlight that elasticity estimation on stiffer tissue with spatio-temporal CNNs is also consistent, while in the literature typically estimated gelatin elasticities in the range of up to 10\% are reported (\cite{kijanka2018local, Engel2015, latus2017}). Naturally, faster shear waves reduce the amount of shear wave information which leads to an increased error for conventional methods. In contrast, stiffer elasticities only lead to slightly reduced performance for our deep learning approach, see Figure~\ref{fig:boxplot_push}. Hence, our findings show that our spatio-temporal CNN approach leads to robust and consistent results across a wide range of elasticities. In general, the performance could be further enhanced by the use of image compounding during data acquisition. The state-of-art is using three angled plane waves for image acquisition which increases the SNR for more robust estimates with ToF \cite{song2012comb}. However, in this case the image acquisition frequency decreases by a factor of three, which results in fewer images containing shear wave information which makes it difficult to estimate shear wave velocity for stiff elasticities. Note that we already observe that performance decreases for stiffer elasticities. Hence, we consider the maximum available imaging frame rate without image compounding.

In this work, we estimate the Young's modulus as a surrogate for tissue elasticity with ground truth annotation performed by indentation experiments. In general, our approach only requires selected material parameters as training targets and subsequently we are able to generalize to unseen data. It is noticeable that predicting known elasticities improves the performance of our spatio-temporal CNN approach. This could be considered as a relevant scenario as further elasticities can be included in the training data. \textcolor{black}{Also, we demonstrate that even in the challenging case of predicting elasticities that are not present during training our spatio-temporal CNN approach leads to competitive performance compared to our ToF method. This demonstrates that our spatio-temporal CNN approach generalizes well between different elasticities even with few ground truth elasticities. Figure~\ref{fig:boxplot_push} suggests a better MAE for ToF for elasticities under 72 kPa, when evaluated elasticities are left out during training. However, ToF-results do not include failed estimates, i.e., we exclude all outlier ToF estimations which are not in the range of 0.1-10m/s. This leads to several missing ToF estimates as shown in Figure~\ref{fig:singelPushPrediction} in gray.} Notably, previous work on estimating the elasticity from SWEI data build on simulated data (\cite{vasconcelos2021viscoelastic,ahmed2021dswe}). However, this only solves the inverse problem of the model underlying the simulation. In addition, performance is limited on real data as, e.g., noise and image artifacts are not sufficient represented in simulated data \cite{ahmed2021dswe}. In contrast, our approach is trained with real data, which includes, e.g., imaging noise or probe artifacts. Also, our local approach can be adapted to real tissue samples, e.g., obtained from tumor resections, by using pathological tumor properties as training targets. Subsequently, pathological properties of soft tissue can be predicted and imaged with our local estimation approach.

\textbf{Push dependency.} Second, we study the robustness of our approach concerning the spatio-temporal window position relative to the push location. Our results, depicted in Figure~\ref{fig:singelPushPrediction} and Figure~\ref{fig:boxplot_push}, show qualitatively and quantitatively that our deep learning approach provides more consistent estimates than ToF and provides accurate estimations independent of the push location relative to the ROI. \textcolor{black}{In the case that all elasticities are present during training our deep learning approach outperforms ToF.} \textcolor{black}{We studied similar result in soft tissue phantoms with consistent predictions for the individual push locations.} This is in contrast to other work in the field that require, e.g., two fine-tuned pushes to the left and right of the ROI \cite{kijanka2018local}. Furthermore, it stands out that with a spatio-temporal CNN we can perform predictions within the push location, where shear wave propagation is complex and diffuse and the imaging is dominated by relaxation dynamics. \textcolor{black}{Predictions can even be performed inside the push region for small spatial window sizes of \SI{5}{} $\times$ \SI{5}{pixels}. It can be assumed that the network learns the relaxation dynamics of the gelatin which changes for different elasticities.} This has not yet been  shown in any previous work that uses deep learning methods in combination with simulated data (\cite{ahmed2021dswe,vasconcelos2021viscoelastic}) or conventional methods \cite{song2012comb}. Still, further investigation on the push depth as well as the material viscosity are necessary. We also study our approach on the complete FOV, while only training with image crops from the ROI. In this way, we evaluate our approach for completely unseen locations relative to the push location and are able to study the performance far away from the push location. For push locations (push  one  and seven) not included in the training data, we can still perform accurate estimates, see Figure~\ref{fig:pushlocation}. This demonstrates that our approach also leads to robust results for unknown push locations. Moreover, our results in Figure~\ref{fig:pushlocation} further confirm that our spatio-temporal CNN approach outperforms ToF and provides accurate estimates for a much larger FOV than the ROI. Similar to ToF, it stands out that our spatio-temporal CNN approach does not provide consistent estimates far away from the push location. This shows that our approach does not over-fit on specific phantom features such as speckle characteristics and fails when no wave information is present in the data. 

\begin{table}
    \caption{MAE for inclusion (in) and background (bg) and the Dice coefficient with the corresponding mean $\mu$ for all five inclusion shapes. Metrics for the spatio-temporal CNN are derived with spatio-temporal windows with a size of 65$\times$65 pixels. The inclusion shapes are displayed in Figure~\ref{Fig_Inclusions} with, e.g., row one referring to phantom $\#1$.}
\label{tab:InclusionMetrics}
\centering
\begin{tabular}{lcccc}
        Method & $\#$ & MAE$_{in}$ [\SI[]{}{\kilo\pascal}]  & MAE$_{bg}$ [\SI[]{}{\kilo\pascal}] & Dice\tabularnewline
        \hline 
        3D CNN  & 1 &\SI[]{10.16(1313)}{}  &\SI[]{0.97(0230)}{}    & \SI[]{0.93}{}    \tabularnewline
            & 2 & \SI[]{19.52(1906)}{}   &\SI[]{1.06(0325)}{}   & \SI[]{0.81}{}    \tabularnewline
            & 3 &  \SI[]{4.70(0832)}{}  &\SI[]{2.42(0501)}{}    & \SI[]{0.98}{}    \tabularnewline
            & 4 &  \SI[]{5.50(1098)}{}  &\SI[]{2.13(0525)}{}    & \SI[]{0.96}{}    \tabularnewline
            & 5 &  \SI[]{6.11(1151)}{}   &\SI[]{2.27(0584)}{}   & \SI[]{0.95}{}    \tabularnewline
            & \textbf{$\mu$}  & \textbf{ \SI[]{7.50(1287)}{}}    & \textbf{\SI[]{1.64(0432)}{}}  & \textbf{\SI[]{0.93}{}  }  \tabularnewline
            \hline
    ToF & 1& \SI[]{15.39(1014)}{}   &   \SI[]{09.19(1009)}{} & \SI[]{0.79}{}\tabularnewline
    & 2& \SI[]{13.28(1038)}{}   &   \SI[]{08.37(89)}{} & \SI[]{0.85}{}\tabularnewline
    & 3& \SI[]{17.14(966)}{}   &   \SI[]{14.49(924)}{} & \SI[]{0.89}{}\tabularnewline
    & 4& \SI[]{13.87(905)}{}   &   \SI[]{13.60(0977)}{} & \SI[]{0.93}{}\tabularnewline
    & 5& \SI[]{19.91(1015)}{}   &   \SI[]{11.99(1089)}{} & \SI[]{0.86}{}\tabularnewline
    & \textbf{$\mu$}  & \textbf{ \SI[]{16.28(1005)}{}}    & \textbf{\SI[]{11.11(1008)}{}}  & \textbf{\SI[]{0.86}{}  } 
\end{tabular}

\end{table}

\begin{figure*}
    \import{./}{fig_inclusionsToF64.tex}
\end{figure*}

\begin{table}
    \caption{Mean MAE and Dice score for all inclusion shapes. Given is the inclusion (in) and background (bg) with all studied spatio-temporal window sizes. The binarization threshold is set to  \SI[]{67.38}{\kilo\pascal}.} 
\label{tab:inclusionMetricsWindowSizes}
\centering
\begin{tabular}{lccc}
    $h_{s} \times w_{s}$ & MAE$_{in}$ [\SI[]{}{\kilo\pascal}] & MAE$_{bg}$ [\SI[]{}{\kilo\pascal}] & Dice \tabularnewline
    \hline 
    $5\times 5$     &  \SI[]{29.68(1384)}{}    & \SI[]{6.37(747)}{}  & \SI[]{0.60}{}   \tabularnewline
    $9\times 9$     &  \SI[]{22.74(1443)}{}    & \SI[]{5.30(754)}{}  & \SI[]{0.75}{}   \tabularnewline
    $17\times 17$   &   \SI[]{14.29(1359)}{}    & \SI[]{4.21(781)}{}  & \SI[]{0.86}{}   \tabularnewline
    $33\times 33$  &   \SI[]{11.45(1355)}{}    & \SI[]{1.91(505)}{}  & \SI[]{0.90}{}  \tabularnewline
    $65\times 65$   &   \SI[]{7.50(1287)}{}    & \SI[]{1.64(432)}{}  & \SI[]{0.93}{}   \tabularnewline
\end{tabular}

\end{table}

\begin{figure}
    \import{./}{fig_diceWindowSizes.tex}
\end{figure}

\begin{figure}
    \import{./}{fig_inclusionChicken.tex}
\end{figure}

\textbf{Inference and Performance.} Third, we study the performance of our network concerning inference time and spatio-temporal window sizes, see Table~\ref{tab:All-networks-with metrics}. Increasing the spatial window size leads to more accurate results compared to using smaller spatial window sizes. This is most likely related to the fact that larger window sizes cover a larger spatial area, hence providing more information about wave propagation. However, using smaller window sizes allows for notably increased model throughput, which is important to provide real-time estimates for larger FOVs and higher resolution. Considering our results in Table~\ref{tab:All-networks-with metrics}, using a smaller spatial window size, e.g., \SI{33}{} $\times$ \SI{33}{pixels} might be a good starting point for further work as there is similar performance compared to \SI{65}{} $\times$ \SI{65}{pixels}, while the throughput is increased by a factor of $3.32$. In general, pixelwise processing is more computationally expensive than an encoder-decoder architecture applied to the entire image at once. However, CNNs are inherently efficient for this task, because computations can be shared across overlapping regions during testing \cite{sermanet2013overfeat}. Similar, a whole image fully convolutional training \cite{long2015fully} could be used to further speed up the training time. We perform patchwise training, which results in higher batch variance and allows to use different augmentation on image crops from the same ROI. Also, a direct advantage of our approach is that sparse estimates can be performed during inference, e.g., only predicting every $n$th pixel. This allows to scale our approach effectively to larger FOVs while maintaining similar inference times. Overall, our results demonstrate that global elasticity maps can be estimated in real-time using our deep learning approach. In particular, the use of more powerful hardware will improve the inference time of our method. Although a comparison due to different hardware is difficult, our spatio-temporal CNN approach is more time efficient than conventional methods and can perform predictions on a smaller window size, e.g., Kijanka et al. \cite{Kijanka2020} report an inference time of \SI{0.22}{ms} per estimate for a spatial window size of \SI{4.5}{} $\times$ \SI{4.5}{\milli\metre} while our spatio-temporal CNN achieves a inference time of \SI{0.07}{ms} for a spatial window size of  \SI{0.32}{} $\times$ \SI{0.4}{\milli\metre}.

\textbf{Inclusion Shapes.} 
Finally, we evaluate our methods on gelatin phantoms with circular stiff inclusions. Our results in Figure~\ref{Fig_Inclusions} demonstrate that our spatio-temporal CNN approach provides consistent estimates with larger spatio-temporal window sizes for the inclusion and the background similar to our results on homogeneous phantoms. Considering the MAE, performance of estimates inside the inclusion increases by a factor of $\sim$ 2 and on the background by a factor of $\sim$ 6 with our spatio-temporal CNN in comparison to ToF. While we perform local estimations, this raises the question how our approach performs on elasticity boundaries with respect to the spatio-temporal window size. Our results in Figure~\ref{Fig_Inclusions} demonstrate that errors can be seen at elasticity boundaries and the shape of the inclusion is still well defined. While smaller window sizes consider a smaller spatial area, this could lead to more distinct boundaries \textcolor{black}{and less blurring. In general, our results show that larger window sizes lead to more consistent estimates, as seen for background predictions in Figure~\ref{Fig_Inclusions}. However, we find that the general performance drop for smaller window sizes outweighs the potential benefit of reduced blurring.} 
Hence, for larger window sizes the boundary is more distinct visible and the Dice score is higher. Nevertheless, it is noticeable that inclusion boundaries can also be retrieved from small windows sizes, e.g., $\sim$ 1 $\times$ \SI[]{1}{\milli\metre} (17 $\times$ 17 pixels). In direct comparison with ToF, the Dice coefficient is similar for a window size of 17 $\times$ 17 pixels (Figure~\ref{fig:diceWindowSizes}). Hence, spatio-temporal window sizes smaller than 17x17 pixel ($\sim$ 1 $\times$ \SI[]{1}{\milli\metre}) are not favorable. \textcolor{black}{Our results in Figure \ref{Fig_InclusionChicken} demonstrate that elasticity estimation in chicken heart tissue is also feasible with our deep learning approach. The investigation of other soft tissue samples in an interesting next step for future work.} Overall, our spatio-temporal CNN approach shows promising results in the estimation of elasticity in inhomogeneous mediums.

\section{Conclusion}
We present 3D spatio-temporal CNNs for local elasticity estimation from real ultrasound shear wave data, which demonstrate increased performance compared to conventional approaches. Our findings show that spatio-temporal CNNs can retrieve local elastic properties from small spatio-temporal windows while being independent of the push location, and demonstrating consistent performance across various elasticities and inhomogenities.  Further work will include in vitro and in vivo experiments of real soft tissues.

\section*{Acknowledgment}
The authors declare that they have no conflict of interest.

\ifCLASSOPTIONcaptionsoff
  \newpage
\fi

\bibliographystyle{IEEEtran}
\bibliography{cas-refs}

\begin{thebibliography}{10}
\providecommand{\url}[1]{#1}
\csname url@samestyle\endcsname
\providecommand{\newblock}{\relax}
\providecommand{\bibinfo}[2]{#2}
\providecommand{\BIBentrySTDinterwordspacing}{\spaceskip=0pt\relax}
\providecommand{\BIBentryALTinterwordstretchfactor}{4}
\providecommand{\BIBentryALTinterwordspacing}{\spaceskip=\fontdimen2\font plus
\BIBentryALTinterwordstretchfactor\fontdimen3\font minus
  \fontdimen4\font\relax}
\providecommand{\BIBforeignlanguage}[2]{{%
\expandafter\ifx\csname l@#1\endcsname\relax
\typeout{** WARNING: IEEEtran.bst: No hyphenation pattern has been}%
\typeout{** loaded for the language `#1'. Using the pattern for}%
\typeout{** the default language instead.}%
\else
\language=\csname l@#1\endcsname
\fi
#2}}
\providecommand{\BIBdecl}{\relax}
\BIBdecl

\bibitem{cosgrove2013efsumb}
D.~Cosgrove \emph{et~al.}, ``{EFSUMB} guidelines and recommendations on the
  clinical use of ultrasound elastography. part 2: Clinical applications,''
  \emph{Eur. J. of Ultrasound}, vol.~34, no.~03, pp. 238--253, 2013.

\bibitem{umale2013experimental}
S.~Umale \emph{et~al.}, ``Experimental mechanical characterization of abdominal
  organs: liver, kidney \& spleen,'' \emph{J Mech. Behav. Biomed. Mater.},
  vol.~17, pp. 22--33, 2013.

\bibitem{yang2017qualitative}
Y.-P. Yang \emph{et~al.}, ``Qualitative and quantitative analysis with a novel
  shear wave speed imaging for differential diagnosis of breast lesions,''
  \emph{Sci. Rep.}, vol.~7, no.~1, pp. 1--11, 2017.

\bibitem{sebag2010shear}
F.~Sebag \emph{et~al.}, ``Shear wave elastography: a new ultrasound imaging
  mode for the differential diagnosis of benign and malignant thyroid
  nodules,'' \emph{J. Clin. Endocrinol. Metab.}, vol.~95, no.~12, pp.
  5281--5288, 2010.

\bibitem{sande2017ultrasound}
J.~A. Sande \emph{et~al.}, ``Ultrasound shear wave elastography and liver
  fibrosis: A prospective multicenter study,'' \emph{World J. Hepatol.},
  vol.~9, no.~1, p.~38, 2017.

\bibitem{bercoff2004supersonic}
J.~Bercoff \emph{et~al.}, ``Supersonic shear imaging: a new technique for soft
  tissue elasticity mapping,'' \emph{IEEE Trans. Ultrason. Ferroelectr. Freq.
  Control}, vol.~51, no.~4, pp. 396--409, 2004.

\bibitem{song2014fast}
P.~Song \emph{et~al.}, ``Fast shear compounding using robust 2-d shear wave
  speed calculation and multi-directional filtering,'' \emph{Ultrasound Med.
  Biol.}, vol.~40, no.~6, pp. 1343--1355, 2014.

\bibitem{latus2017}
S.~Latus \emph{et~al.}, ``An approach for needle based optical coherence
  elastography measurements,'' in \emph{MICCAI}.\hskip 1em plus 0.5em minus
  0.4em\relax Springer, 2017, pp. 655--663.

\bibitem{song2012comb}
P.~Song \emph{et~al.}, ``Comb-push ultrasound shear elastography (cuse): a
  novel method for two-dimensional shear elasticity imaging of soft tissues,''
  \emph{IEEE Trans. Med. Imaging}, vol.~31, no.~9, pp. 1821--1832, 2012.

\bibitem{carrascal2017improved}
C.~A. Carrascal \emph{et~al.}, ``Improved shear wave group velocity estimation
  method based on spatiotemporal peak and thresholding motion search,''
  \emph{IEEE Trans. Ultrason. Ferroelectr. Freq. Control}, vol.~64, no.~4, pp.
  660--668, 2017.

\bibitem{Engel2015}
A.~J. Engel and G.~R. Bashford, ``{A new method for shear wave speed estimation
  in shear wave elastography},'' \emph{IEEE Trans. Ultrason. Ferroelectr. Freq.
  Control}, vol.~62, no.~12, pp. 2106--2114, 2015.

\bibitem{tanter2008quantitative}
M.~Tanter \emph{et~al.}, ``Quantitative assessment of breast lesion
  viscoelasticity: initial clinical results using supersonic shear imaging,''
  \emph{Ultrasound Med. Biol.}, vol.~34, no.~9, pp. 1373--1386, 2008.

\bibitem{yang2014comparative}
J.~Yang \emph{et~al.}, ``Comparative study on shear wave speed estimation
  algorithms in {ARFI} for improving its reliability,'' in \emph{2014 36th
  Annual International Conf. Proc. IEEE Eng. Med. Biol. Soc.}\hskip 1em plus
  0.5em minus 0.4em\relax IEEE, 2014, pp. 226--229.

\bibitem{wang2013precision}
M.~Wang \emph{et~al.}, ``On the precision of time-of-flight shear wave speed
  estimation in homogeneous soft solids: initial results using a matrix array
  transducer,'' \emph{IEEE Trans. Ultrason. Ferroelectr. Freq. Control},
  vol.~60, no.~4, pp. 758--770, 2013.

\bibitem{kijanka2018local}
P.~Kijanka and M.~W. Urban, ``Local phase velocity based imaging: A new
  technique used for ultrasound shear wave elastography,'' \emph{IEEE Trans.
  Med. Imaging}, vol.~38, no.~4, pp. 894--908, 2018.

\bibitem{Kijanka2020}
P.~Kijanka and M.~Urban, ``Fast local phase velocity-based imaging: Shear wave
  particle velocity and displacement motion study,'' \emph{IEEE Trans.
  Ultrason. Ferroelectr. Freq. Control}, vol.~67, no.~3, pp. 526--537, 2019.

\bibitem{kibria2018gluenet}
M.~Kibria \emph{et~al.}, ``Gluenet: Ultrasound elastography using convolutional
  neural network,'' in \emph{Simulation, Image Processing, and Ultrasound
  Systems for Assisted Diagnosis and Navigation}.\hskip 1em plus 0.5em minus
  0.4em\relax Springer, 2018, pp. 21--28.

\bibitem{chan2021deep}
D.~Y. Chan \emph{et~al.}, ``Deep convolutional neural networks for displacement
  estimation in arfi imaging,'' \emph{IEEE Trans. Ultrason. Ferroelectr. Freq.
  Control}, vol.~68, no.~7, pp. 2472--2481, 2021.

\bibitem{kz2020semi}
A.~KZ~Tehrani \emph{et~al.}, ``Semi-supervised training of optical flow
  convolutional neural networks in ultrasound elastography,'' in
  \emph{MICCAI}.\hskip 1em plus 0.5em minus 0.4em\relax Springer, 2020, pp.
  504--513.

\bibitem{delaunay2021unsupervised}
R.~Delaunay \emph{et~al.}, ``An unsupervised learning approach to ultrasound
  strain elastography with spatio-temporal consistency,'' \emph{Phys. Med.
  Biol.}, vol.~66, no.~17, p. 175031, 2021.

\bibitem{jin2021deep}
F.~Q. Jin \emph{et~al.}, ``Deep learning based quantitative uncertainty
  estimation for ultrasound shear wave elasticity imaging,'' in \emph{2021 IEEE
  Int. Ultrason. Symp. (IUS)}.\hskip 1em plus 0.5em minus 0.4em\relax IEEE,
  2021, pp. 1--4.

\bibitem{vasconcelos2021viscoelastic}
L.~Vasconcelos \emph{et~al.}, ``Viscoelastic parameter estimation using
  simulated shear wave motion and convolutional neural networks,''
  \emph{Computers in Biology and Medicine}, vol. 133, p. 104382, 2021.

\bibitem{ahmed2021dswe}
S.~Ahmed \emph{et~al.}, ``Dswe-net: A deep learning approach for shear wave
  elastography and lesion segmentation using single push acoustic radiation
  force,'' \emph{Ultrasonics}, vol. 110, p. 106283, 2021.

\bibitem{tran2015learning}
D.~Tran \emph{et~al.}, ``Learning spatiotemporal features with 3d convolutional
  networks,'' in \emph{Proc. of the ICCV}, 2015, pp. 4489--4497.

\bibitem{neidhardt2020deep}
M.~Neidhardt \emph{et~al.}, ``Deep learning for high speed optical coherence
  elastography,'' in \emph{2020 IEEE 17th Int. Symposium Biomed. Imaging
  (ISBI)}.\hskip 1em plus 0.5em minus 0.4em\relax IEEE, 2020, pp. 1583--1586.

\bibitem{neidhardt20214d}
Neidhardt \emph{et~al.}, ``4d deep learning for real-time volumetric optical
  coherence elastography,'' \emph{Int. J. Comput. Assist. Radiol. Surg.},
  vol.~16, no.~1, pp. 23--27, 2021.

\bibitem{huang2017densely}
G.~Huang \emph{et~al.}, ``Densely connected convolutional networks,'' in
  \emph{Proc. of the IEEE conference on CVPR}, 2017, pp. 4700--4708.

\bibitem{batchnorm2015}
S.~Ioffe and C.~Szegedy, ``Batch normalization: Accelerating deep network
  training by reducing internal covariate shift,'' \emph{arXiv preprint
  arXiv:1502.03167}, 2015.

\bibitem{sarvazyan2013acoustic}
A.~P. Sarvazyan \emph{et~al.}, ``Acoustic waves in medical imaging and
  diagnostics,'' \emph{Ultrasound Med. Biol.}, vol.~39, no.~7, pp. 1133--1146,
  2013.

\bibitem{Samani2003}
A.~Samani \emph{et~al.}, ``Measuring the elastic modulus of ex vivo small
  tissue samples,'' \emph{Phys. Med. Biol.}, vol.~48, no.~14, p. 2183, 2003.

\bibitem{arda2011quantitative}
K.~Arda \emph{et~al.}, ``Quantitative assessment of normal soft-tissue
  elasticity using shear-wave ultrasound elastography,'' \emph{AJR Am. J.
  Roentgenol.}, vol. 197, no.~3, pp. 532--536, 2011.

\bibitem{samur2007robotic}
E.~Samur \emph{et~al.}, ``A robotic indenter for minimally invasive measurement
  and characterization of soft tissue response,'' \emph{Med. Image Anal.},
  vol.~11, no.~4, pp. 361--373, 2007.

\bibitem{kot2012elastic}
B.~C.~W. Kot \emph{et~al.}, ``Elastic modulus of muscle and tendon with shear
  wave ultrasound elastography: variations with different technical settings,''
  \emph{PloS one}, vol.~7, no.~8, p. e44348, 2012.

\bibitem{ahn2010mechanical}
B.-M. Ahn \emph{et~al.}, ``Mechanical property characterization of prostate
  cancer using a minimally motorized indenter in an ex vivo indentation
  experiment,'' \emph{Urology}, vol.~76, no.~4, pp. 1007--1011, 2010.

\bibitem{krouskop1998elastic}
T.~A. Krouskop \emph{et~al.}, ``Elastic moduli of breast and prostate tissues
  under compression,'' \emph{Ultrason. Imaging}, vol.~20, no.~4, pp. 260--274,
  1998.

\bibitem{samani2007elastic}
A.~Samani \emph{et~al.}, ``Elastic moduli of normal and pathological human
  breast tissues: an inversion-technique-based investigation of 169 samples,''
  \emph{Phys. Med. Biol.}, vol.~52, no.~6, p. 1565, 2007.

\bibitem{forte2015modelling}
A.~Forte \emph{et~al.}, ``Modelling and experimental characterisation of the
  rate dependent fracture properties of gelatine gels,'' \emph{Food
  Hydrocoll.}, vol.~46, pp. 180--190, 2015.

\bibitem{delaine2016experimental}
R.~Delaine-Smith \emph{et~al.}, ``Experimental validation of a flat punch
  indentation methodology calibrated against unconfined compression tests for
  determination of soft tissue biomechanics,'' \emph{J Mech. Behav. Biomed.
  Mater.}, vol.~60, pp. 401--415, 2016.

\bibitem{zmudzinska2018assessment}
M.~{\.Z}mudzi{\'n}ska \emph{et~al.}, ``The assessment of the applicability of
  shear wave elastography in modelling of the mechanical parameters of the
  liver,'' \emph{Acta Bioeng. Biomech.}, vol.~20, no.~4, 2018.

\bibitem{loupas1995axial}
T.~Loupas \emph{et~al.}, ``An axial velocity estimator for ultrasound blood
  flow imaging, based on a full evaluation of the doppler equation by means of
  a telastic moduli of normal and pathological human breast tissues: an
  inversion-technique-based investigation of 169 sampleswo-dimensional
  autocorrelation approach,'' \emph{IEEE Trans. Ultrason. Ferroelectr. Freq.
  Control}, vol.~42, no.~4, pp. 672--688, 1995.

\bibitem{Trutna2020}
C.~A. Trutna \emph{et~al.}, ``{Measurement of Viscoelastic Material Model
  Parameters Using Fractional Derivative Group Shear Wave Speeds in Simulation
  and Phantom Data},'' \emph{IEEE Trans. Ultrason. Ferroelectr. Freq. Control},
  vol.~67, no.~2, pp. 286--295, 2020.

\bibitem{sermanet2013overfeat}
P.~Sermanet \emph{et~al.}, ``Overfeat: Integrated recognition, localization and
  detection using convolutional networks,'' \emph{arXiv preprint
  arXiv:1312.6229}, 2013.

\bibitem{long2015fully}
J.~Long \emph{et~al.}, ``Fully convolutional networks for semantic
  segmentation,'' in \emph{Proc. of the IEEE conference on CVPR}, 2015, pp.
  3431--3440.

\end{thebibliography}

\end{document}